\documentclass[usenatbib,useAMS,a4paper]{mn2e}

\usepackage{txfonts}
\usepackage{epsfig}

\def\aaps{{AAPS}}
\def\mnras{{MNRAS}}
\def\apj{{ApJ}}
\def\apjl{{ApJL}}
\def\apjs{{ApJS}}

\def\aap{{As\&A}}
\def\aj{{AJ}}

\def\msun{{\rm M_{\odot}}}

\title [On the lifetime of discs around late type stars]{On the lifetime of discs around late type stars}
\author[]{Barbara Ercolano$^{1,2}$, Nate Bastian$^{1}$, Loredana
  Spezzi$^{3}$, and James Owen$^{4}$\\
$^1$ Excellence Cluster Universe, Boltzmannstr. 2, 85748 Garching, Germany\\
$^2$ University Observatory, Ludwig-Maximillians University Munich, Scheinerstr. 1, 81679 Munich, Germany\\
$^3$ Research and Scientific Support Department, European Space Agency (ESA-ESTEC), P.O. Box 299, 2200 AG Noordwijk, The
Netherlands\\
$^4$ Institute of Astronomy, Madingley Road, Cambridge CB3 0HA, UK\\
}

\date{Submitted: }

\pagerange{\pageref{firstpage}--\pageref{lastpage}} \pubyear{2003}

\begin{document}
\def\lta{\mathrel{\spose{\lower 3pt\hbox{$\mathchar"218$}}
     \raise 2.0pt\hbox{$\mathchar"13C$}}}
\def\gta{\mathrel{\spose{\lower 3pt\hbox{$\mathchar"218$}}
     \raise 2.0pt\hbox{$\mathchar"13E$}}}
\def\Msun{{\rm M}_\odot}
\def\msun{{\rm M}_\odot}
\def\Rsun{{\rm R}_\odot}
\def\Lsun{{\rm L}_\odot}
\def\19{GRS~1915+105}
\label{firstpage}
\maketitle

\begin{abstract}

We address the question of whether protoplanetary discs around low
mass stars (e.g. M-dwarfs) may be longer lived than their solar-type
counterparts. This question is particularly relevant to assess the
planet-making potential of these lower mass discs. Given the
uncertainties inherent to age-dating young stars, we propose an
alternative approach that is to analyse the spatial distribution of
disc-bearing low-mass stars and compare it to that of  disc-bearing
solar-type stars in the same cluster. A significant age difference
between the two populations should be reflected in their average
nearest neighbour distance (normalised to the number of sources),
where the older population should appear more spread out.  

To this aim, we perform a Minimum Spanning Tree (MST) analysis on the
spatial distribution of disc-bearing young stellar objects (YSOs) in
six nearby low mass star forming regions.  
We find no evidence for significant age differences between the
disc-bearing low-mass (later than M2) and 'solar-type' (earlier than M2) stars in these
regions. We model our results by constructing and analysing synthetic
fractal distributions that we evolve for typical values of the velocity
dispersions. A comparison of simple models to our MST analysis
suggests that the lifetime of discs around M-stars is similar to that
of discs around solar-type stars. Furthermore, a model-independent
spatial analysis of the observations robustly shows that any age
differences between the two samples must be smaller than the average
age difference between disc-bearing classical T-Tauri stars and
disc-less Weak-Lined T-Tauri stars. 

\end{abstract}


\section{Introduction}
The lifetime of protoplanetary discs around young low-mass stars has
been the subject of numerous recent observational and theoretical
studies (e.g. Luhman et al 2010; Ercolano, Clarke \& Hall 2011; Currie \& Kenyon 2009; Ercolano, Clarke \& Robitaille 2009; Sicilia-Aguilar et al 2008). The interest is justified by the key role played by
this circumstellar material in the formation and evolution of
planetary systems, by providing the gas and dust reservoirs from which
planets form and through which they migrate. 

Discs around solar type stars are rarely seen for systems of age 10
Myr or older (e.g. Mamejek, 2009), implying that (giant) planet formation around such stars
must occur within this timescale. There has been considerable
interest recently with regards to planetary systems around lower
mass stars (e.g. Pascucci et al 2011). The attraction of M-dwarfs
resides, first of all, in their larger
population, assuming a Salpeter/Kroupa initial mass function
(Salpeter 1955; Kroupa 2002) there are 50/10
0.2~M$_{\odot}$ stars for every 1~M$_{\odot}$ star. Furthermore, 
their habitable zones (HZ) extend close to the parent star allowing for a
stronger (detectable) radial velocity signal from potentially
neptune-size planets or smaller. 

There are, however, a number of possible drawbacks to the search of
planetary systems around M-dwarfs, namely their stronger magnetic
activity and the possibility of tidal locking for planets with small orbital
separation from the star. There is at present no consensus
with regards to whether either of these factors could prevent the
development of life on these planets (for a recent review see Pascucci et
al 2011). With
regards to the actual formation of the planets themselves there are further
considerations to be made. The lower column densities of discs around
late type stars may imply 
longer timescales for planet formation under the core accretion scenario, but this may be offset
by the fact that the discs may be longer lived. While absolute ages of star forming regions (or, worse,
individual objects within a region) are known to be uncertain
(e.g. Mayne \& Naylor~2008; Baraffe, Chabrier, \& Gallardo~2009), the 
high fraction of disc bearing M-stars in $\eta$ Chamaleontis, aged 8
Myr, has been interpreted as
evidence for longer disc lifetimes around late type stars
(Sicilia-Aguilar et al 2009).  Furthermore, there is tentative evidence
that the frequency of infrared
excess emission appears to decrease less steeply with age as the 
mass of the central object decreases, this result, however, is not yet
conclusive because only a few disk fractions have been measured for low-mass
stars older than 3 Myr (Carpenter et al. 2006; Luhman 2009).

A significantly longer disc lifetime for late-type stars
could have important implications for the 
formation and evolution of planetary systems, but compelling
observational evidence is still missing. Theoretical considerations
based on standard 
disc dispersal mechanisms do not seem to support a significantly
longer disc lifetime for M-stars.  
Gorti, Dullemond \& Hollenbach (2009) calculated
models of disc dispersal driven mainly by FUV-driven photoevaporation
for stars in the range 0.5-30M$_{\odot}$, finding that disc lifetimes
were only mildly dependent on stellar mass in the range
0.5-3M$_{\odot}$, with roughly only a factor two difference in the 
timescales at 0.5 and 3M$_{\odot}$. It is unclear, however, how these
results could be extrapolated down to the M-dwarf regime. 
The X-ray photoevaporation models of Ercolano et al (2008, 2009) and 
Owen et al. (2010) only considered
the dispersal of discs around a 0.7~M$_{\odot}$ star. However, if a
constant viscous $\alpha$ parameter and radial scaling
(i.e. $\nu\propto R$) is  assumed, the viscous scaling relations
(Lynden-Bell \& Pringle, 1974) can be used to estimate median disc
lifetimes as a function of stellar mass. By approximating the disc's
lifetime as the time when the viscous accretion rates become equal to
the wind mass loss rates, described in Owen et al. (2011) as the
`null' model,  (see also Ercolano \& Clarke 2010), we find that:   
\begin{equation} 
\tau_d \propto L_X^{-2/3}t_{\nu}^{1/3}M_d(0)^{2/3} \propto L_X^{-2/3}R_1^{1/3}M_*^{-1/6}M_d(0)^{2/3} 
\end{equation} 
\noindent where the second relation is for a flared reprocessing disc
(e.g. Chiang \& Goldreich, 1997), $t_{\nu}$ is the viscous time, $R_1$
is the initial scale size of the disc and $M_d(0)$ is the initial disc
mass which is expected to scale linearly with mass (e.g. Alexander \&
Armitage, 2006). The X-ray luminosity has been observationally shown
to scale approximately as $L_X\propto M_*^{3/2}$ (e.g. Preibisch et
al. 2005, G\"{u}del et al. 2007). However, the scaling of $t_{\nu}$,
and hence $R_1$, with mass is rather more uncertain, both
observationally and theoretically.  Alexander \& Armitage (2006) use
the observed $\dot{M} \propto M_*^2$ scaling  
\citep[e.g.,][]{Muz05,Nat06,Sic06} to argue that $t_\nu\propto
M_*^{-1}$ which yields $\tau_d \propto M_*^{-2/3}$. However, the
solution for $t_\nu$ from the  $M$--$\dot{M}$ relation is far from
unique (e.g. Dullemond et al. 2006). Furthermore, the observationally
determined $M$--$\dot{M}$ relation has not yet been confirmed as a
physical relation and the role played by observational biases in its
derivation is still uncertain (e.g. Clarke \& Pringle, 2006; Tilling
et al. 2008).  
By instead assuming that all stars form from cores with identical
ratios of rotational to gravitational energy ($\beta$), then
$R_1\propto M_*$ and thus $\tau_d \propto M_*^{-1/6}$. By including
all 
    known scalings in equation (1) one finds:
\begin{equation}
\tau_d\propto M_*^{-1/2}R_1^{1/3}
\end{equation}
Therefore, while the actual variation of disc lifetime with mass is unclear, for all sensible scalings of disc radius with mass, it is clear that X-ray photoevaporation predicts only a mild negative
scaling between disc lifetime and mass.

There are clearly a number of, potentially important, theoretical
uncertainties and these highlight the need to find alternative 
methods, based on observations, to understand the dependence of disc
lifetimes on stellar mass, which is essential to constrain models of
disc evolution and planet formation around M-dwarfs. In this paper we
compare the spatial distribution of disc-bearing young late type stars
(later than M2) to that of earlier types, which we
crudely refer to as `solar-type stars'
(i.e. earlier than M2). The aim is to look for the statistical signature of a
longer lived disc population amongst the lower mass stars, which
should manifest itself as larger mean separations between disc-bearing
low mass stars.  This is based on the simple assumption that the older stars
would have had more time to move away from their birthplace at the
typical velocity dispersions measured for low mass star forming
regions (of order 1 km/s). Indeed a spatial comparison of Class~I and
Class~II sources showed that Class~I sources were less distributed
than the Class~IIs, and were more likely to still be near filamentary
structures in the gas where they presumably formed (e.g. Gutermuth et
al.~2008,  Maaskant et al. 2011 ). 

In Section 2 we describe the hypothesis in more detail as well as the
methods and observational data-sets employed. In Section 3 we present
the results of the analysis, as well as Monte Carlo simulations to
assess the robustness of our results. Finally, a discussion of the physical
implications in given in Section~4. 

\section{Hypothesis, Methods and Observations}

\begin{table}
\begin{center}
\begin{tabular}{lcccc}
\hline
Region        & Age                &  Distance             & Size & $\sigma_v$ \\
                    & [Myr]                &    [pc]                      &  [pc] &   [km/s]      \\  
Serpens   & 2-6$^{1a}$            &    260-415$^{1b}$            &   0.4  & 0.25-0.611$^2$ \\
Lupus III   & 2-6$^{3a,3b}$  &  200$^{3a}$         & 2.4      & 1.3$^4$ \\
Taurus     & 1$^{5a}$                & 140$^{5b}$             & 18    & 0.2$^6$ \\
IC 348      & 2-3$^{7a}$           & 261-340$^{7b}$            & 0.95       & 0.1-0.2$^8$ \\
Cha I        & 2$^{9a}$          & 162$^{9b}$        & 2.8 & 0.6-1.2$^{10}$ \\
ChaII        & 4-5$^{11}$       & 178$^{9b}$        & 2.3   &   -- \\
Tr 37        & 4$^{12a}$        & 900$^{12b}$            & 4.7 & -- \\
Tr 37 West   & 1$^{12a}$        & 900$^{12b}$            & 4.7 & -- \\

\hline
\end{tabular}
\caption{Physical properties of the star forming region [$^{1a}$ \citet{Oli09}; 
$^{1b}$ \citet{Str96} and \citet{Dzi10}; $^2$ Williams 2001; $^{3a}$\citet{Com08}; $^{3b}$ \citet{Mer08};
$^4$Makarov 2007; $^{5a}$ \citet{Luh04,Ken08} and references therein; $^{5b}$ \citet{Ken94}; $^6$ Kraus \& Hillenbrand 2008;
$^{7a}$Muench et al (2003); \citet{Luh03}; $^{7b}$\citet{Her08,Luh03,Sch99,Her98,Cer93}; $^8$Herbig 1998;
$^{9a}$ \citet{Luh07,Luh08}; $^{9b}$ \citet{Whi97}; $^{10}$Dubath et al. 1996; $^{11}$\citet{Spe08};  $^{12a}$ \citet{Sic05,Sic06}; $^{12b}$ \citet{Con02}.}
\end{center}
\end{table}

\begin{table}
\begin{center}
\begin{tabular}{lcccc}
\hline
Region    & \multicolumn{2}{|c|}{Selected Number}   & $\Delta$d$_{av}$ & 3$\times$Error \\
               & $<$M2 & $>$M2 & [pc] & [pc]    \\  
Serpens      & 12   & 26    & +0.014 & 0.075 \\ 
Lupus III    &  28    & 22  & -0.132    & 0.033 \\
Taurus       & 104  & 102 & -0.115    & 0.110 \\
IC 348        &  65  & 26   & +0.002 & 0.012 \\ 
Cha I          & 61    & 32  & -0.004  & 0.054\\
ChaII          & 21    & 18   & -0.076  & 0.089    \\
\hline
\end{tabular}
\caption{Observed average separation differences,
  $\Delta$d$_{av} = $d$_{av}(<M2)$-d$_{av}(>M2)$ , between
  disc-bearing low mass and solar-mass stars in young clusters. }
\end{center}
\end{table}

If it is true that discs around low mass stars live significantly
longer than discs around more massive (solar-type) stars, and that
stars in a cluster are formed over an extended period (i.e. the
age-spread within a cluster is comparable to the timescale over which
protoplanetary discs disperse), then the disc-bearing low-mass
(later than M2)  stellar
population in a given cluster should be on average older
than the disc-bearing solar-mass population (earlier than M2) in
the same cluster. The mean age difference amongst the two stellar
populations should be reflected in the relative spatial
distribution of the young stars, with the disc-bearing low mass stars
being more spread out than the solar-mass stars. Indeed at a velocity
of 1 km/s, which is a typical velocity dispersion measured for nearby low mass
star-forming region, a disc-bearing M-star would be able to travel
approximately 10~pc during a hypothetical 10-Myr lifetime of its
disc, while a disc-bearing solar-type star may only move a fraction of
that distance. 

The recent compilation of catalogues of young stellar objects (YSOs) in
nearby star-forming regions renders possible a preliminary
investigation of the spatial distribution of disc-bearing YSOs.  We
have collected data for seven nearby star forming
regions, including the position of the YSOs, their spectral classification
and whether an infra-red excess (i.e. a disc) is detected for each
individual object. In Table~1 we summarise the physical properties of
the regions studied and provide references for the data we used. We
note that the `size' quoted in the Table is not the size commonly
given in the literature for these regions but it is an effective
radius of the region occupied by the well characterised, disc-bearing
YSOs used in our analysis. This was obtained by simply plotting the
disc bearing sources, calculating the area occupied by them and
deriving an effective radius from it.

\subsection{Spatial analysis}
We employ a Minimum-Spanning-Tree (MST) method to calculate average
stellar separations. A MST is formed by connecting all points (spatial
positions in this case) in order to form a unified network, such that
the total length (i.e. sum) of all of the connections is minimised,
and no closed loops are formed. The method is routinely employed to study
large scale star-forming regions in galaxies (e.g. Bastian et
al. 2007, 2009, 2011; Gieles et al. 2008; Schmeja et al 2009) as well as local star-forming
regions (Koenig et al. 2008; Gutermuth et al. 2009; Schmeja, Kumar \&
Ferreira 2008). We constructed
MST diagrams for the disc-bearing low-mass and solar-type stars in each
region and calculated the average nearest-neighbour distance ($d_{av}$)
for the two populations. We ensured that the two populations had the
same number of sources by stochastically culling the number of
sources in the larger population. The stochastic culling and nearest
neighbours analysis of the larger sample was performed fifty times and
an average value was taken. The spread in the Monte Carlo results
provided us with a measure of the error introduced by this procedure
and by the low number statistics. 
We only included YSOs in the sample for which spectral types were available from the literature and excluded brown dwarfs from our samples.
Spectral types were measured on the basis of optical and/or near-infrared follow-up spectroscopy. 
For a more detailed description of the spectral type classification of the YSOs in our sample, we defer the reader to the work by 
\citet{Oli09} for Serpens, \citet{Hug93,Kra97,Com03,All07} for Lup~III,  \citet{Luh04} and references therein for Taurus, \citet{Luh03} and Muench et al (2003) for IC~348, 
\citet{Luh07} and  \citet{Luh08b} for Cha~I, \citet{Spe08} for Cha~II and \citet{Sic05} for Tr~37.
As will be further discussed in Section 3, we find no evidence of
longer disc dispersal timescales around the lower mass stellar
population.
Figure 1 shows the distribution of the disc-bearing sources as a function of spectral types in the range G6 to M6\footnote{Note that brown dwarf were excluded from our sample, however due to their small numbers, their inclusion does not affect the conclusions},  in all the clusters studied, apart from Tr~37, which, as described in the following section, has been excluded from our final analysis. The thick horizontal line shows the formal separation between "solar types" and "low mass" stars set at M2 in this work. 

 \begin{figure*}
 \begin{center}
 \includegraphics[width=8cm]{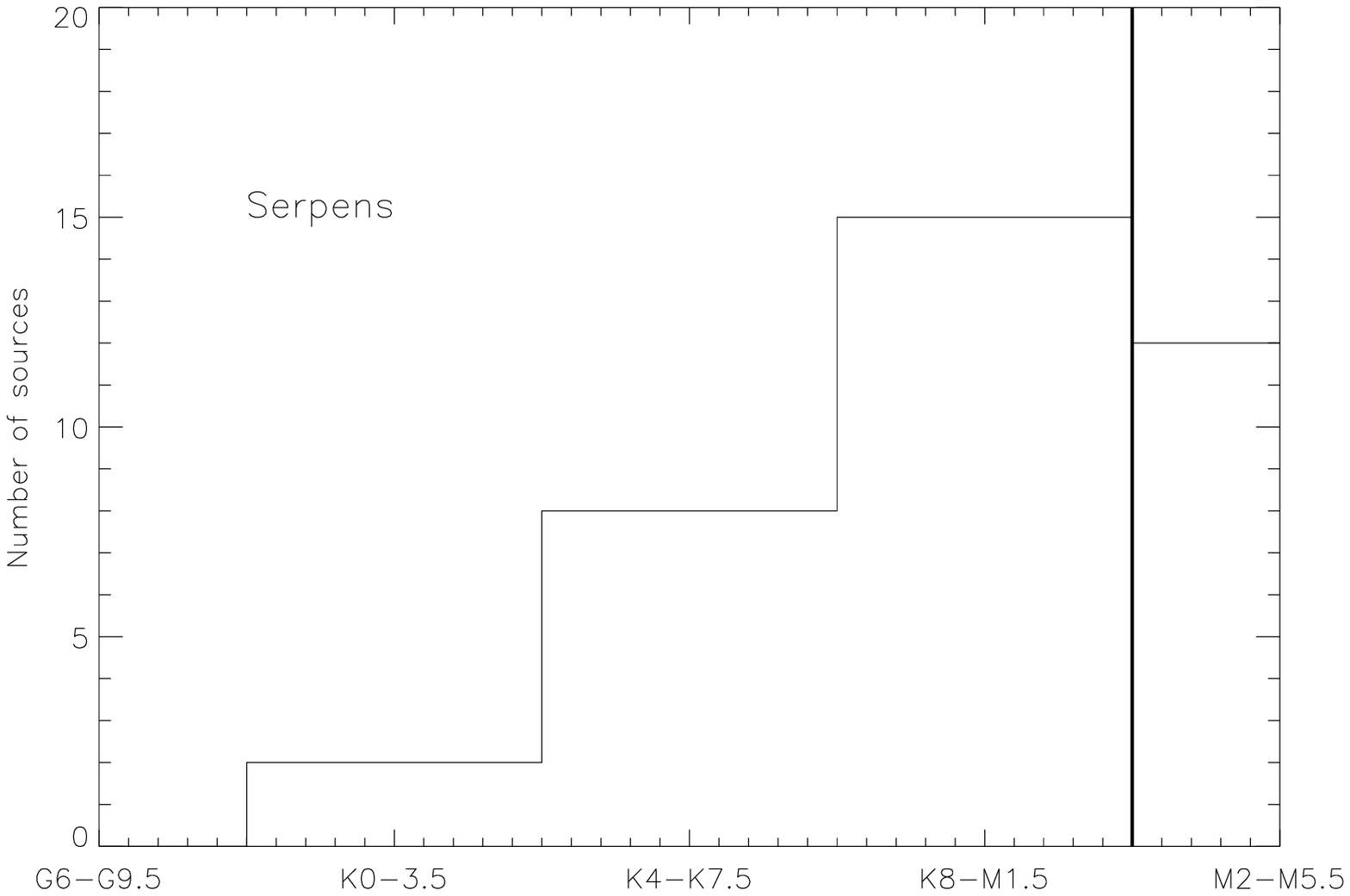}
 \includegraphics[width=8cm]{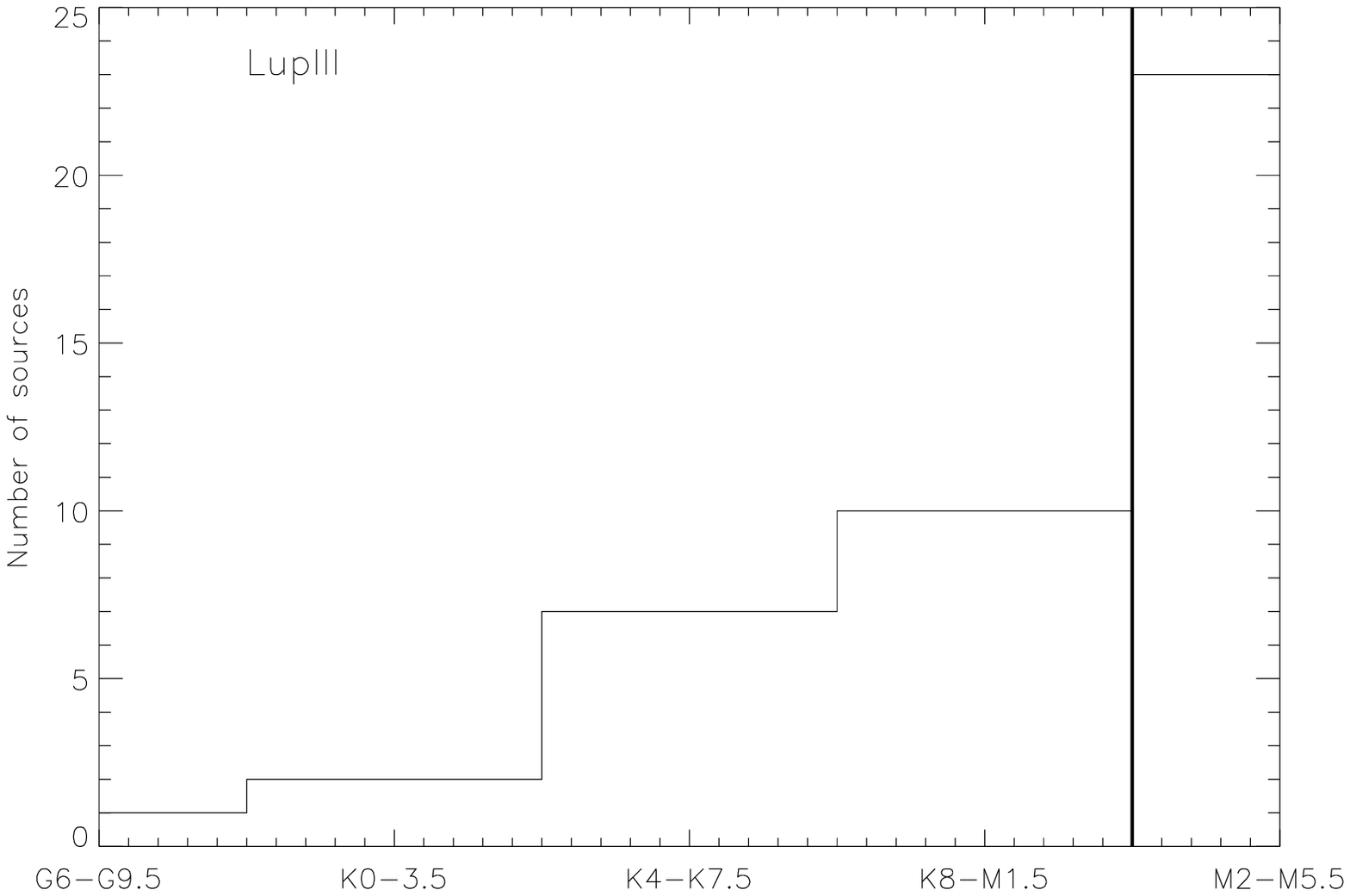}
 \includegraphics[width=8cm]{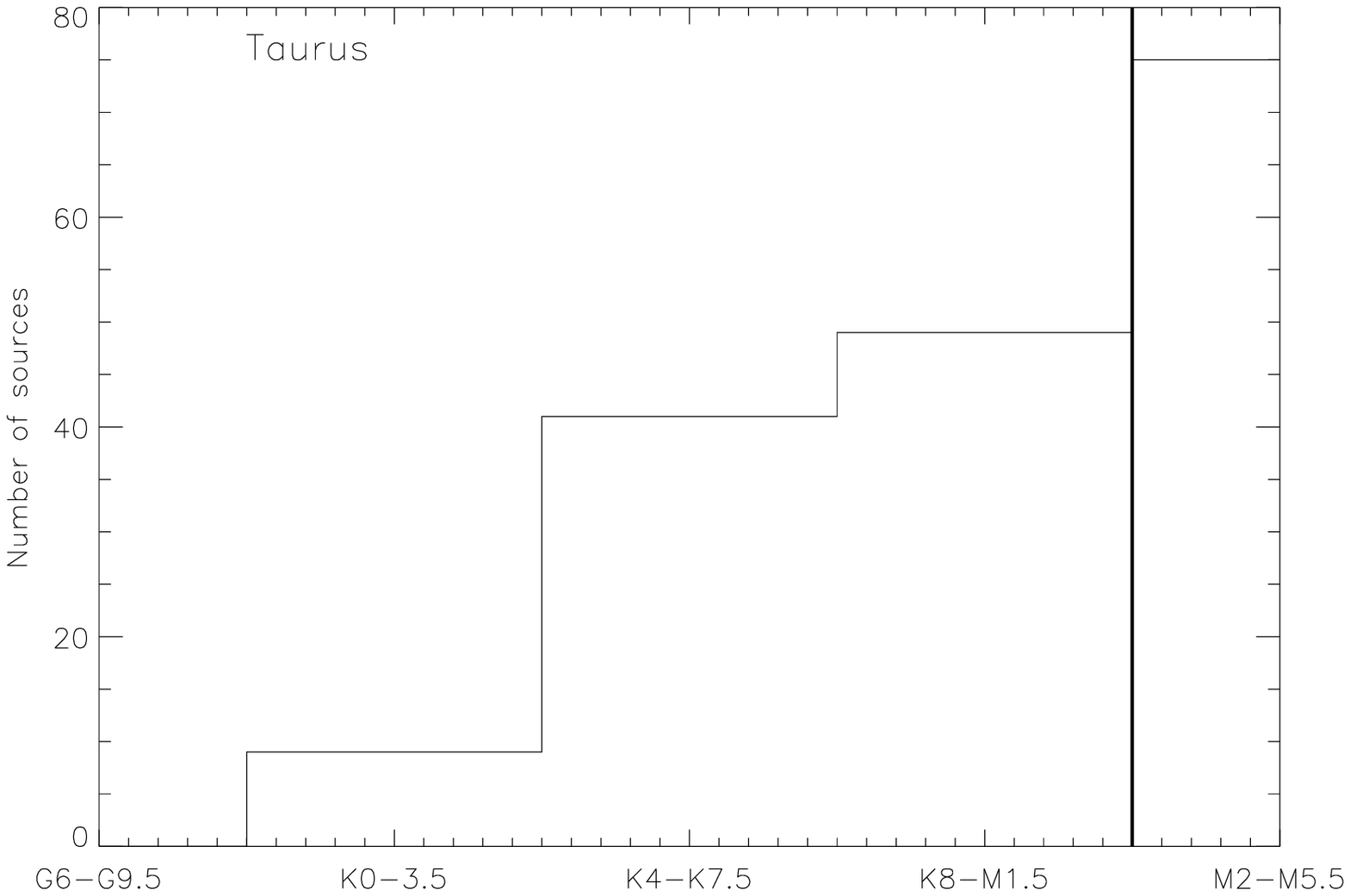}
 \includegraphics[width=8cm]{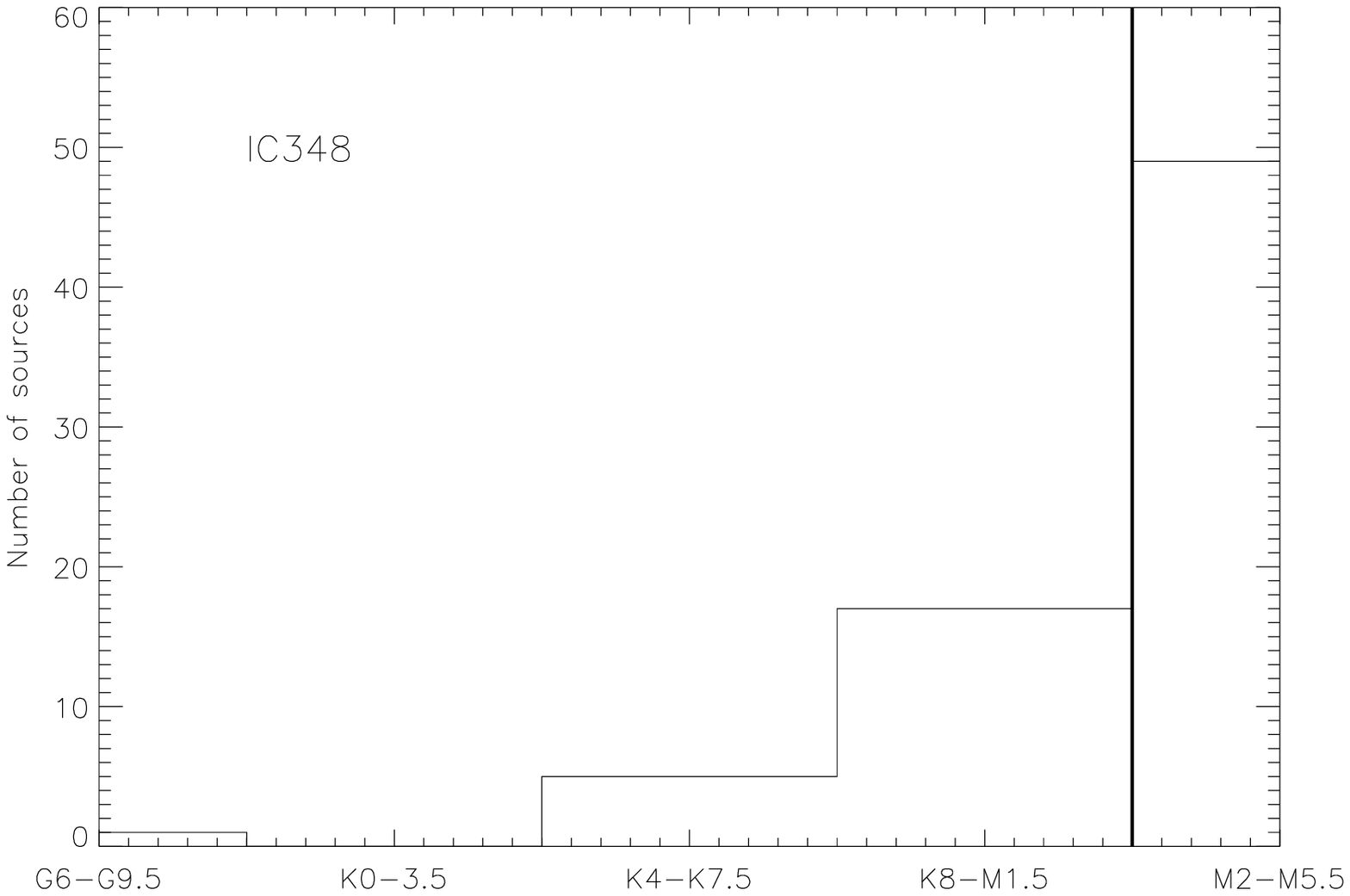}
 \includegraphics[width=8cm]{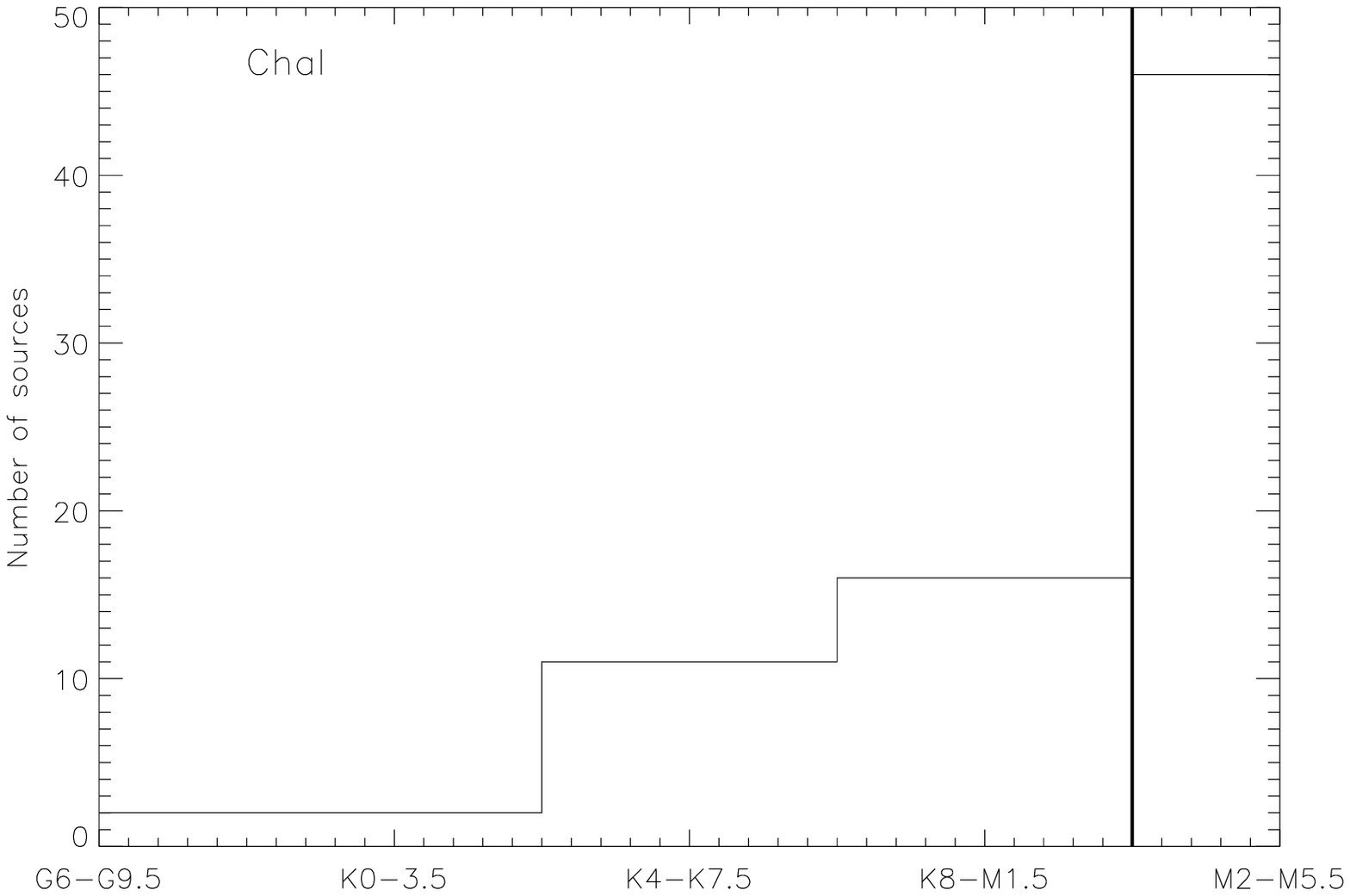}
 \includegraphics[width=8cm]{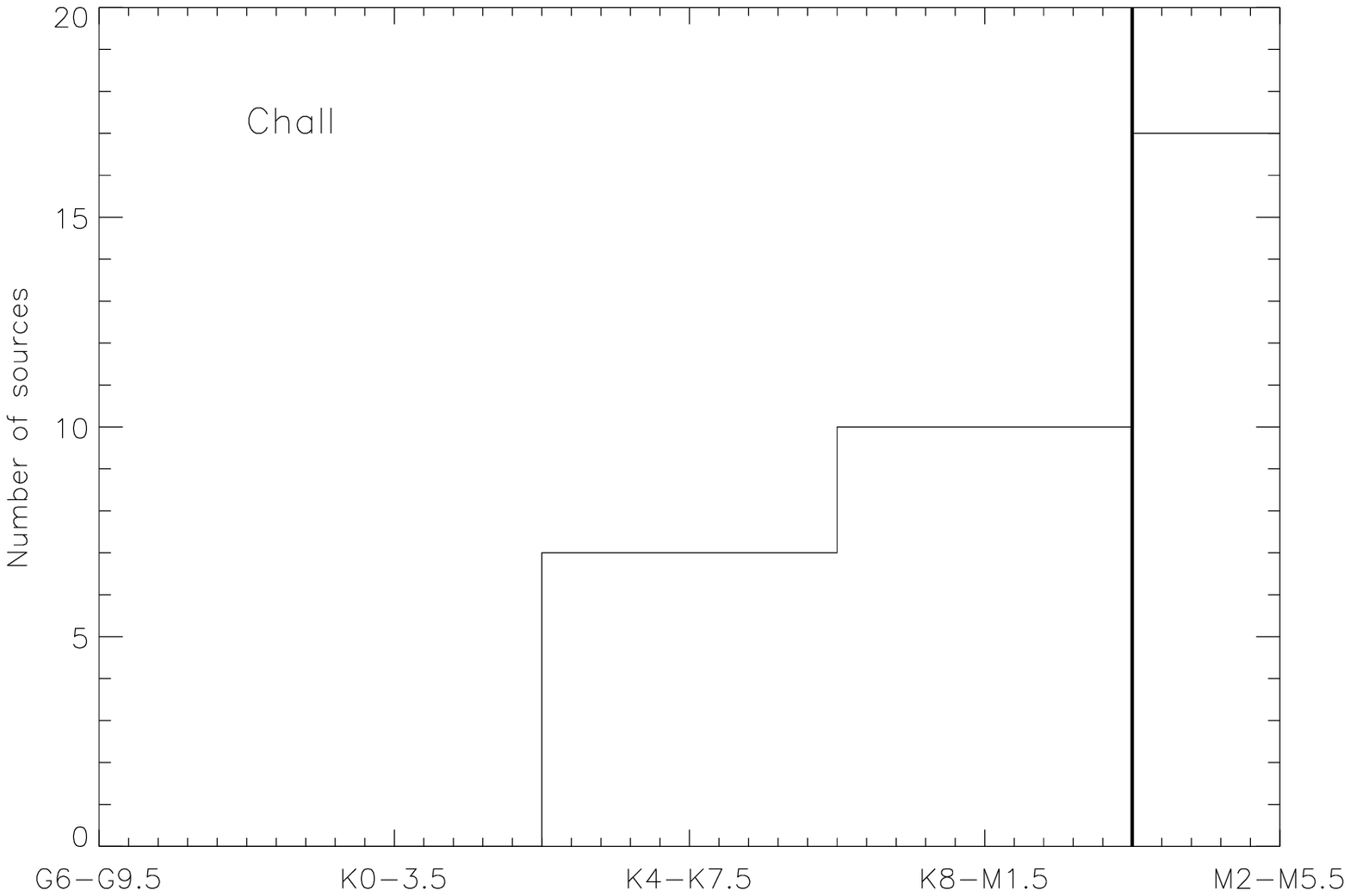}
\caption{The distribution of spectral types of the disc-bearing sources in the region studied. The thick horizontal line shows the formal separation between "solar types" and "low mass" stars set at M2 in this work.} 
 \end{center}
 \end{figure*}

\subsection{Method validation}
As is often the case in this field, however, our analysis
is plagued by small number statistics and perhaps a signal that may be
too small to be detected using the proposed technique. In order to
validate our method, we perform two sets of tests which are described
next. 

\subsubsection{Test 1: The spatial distribution of disc-bearing and
  disc-less objects}

The average age difference between the disc-bearing (class II) and
disc-less (class III) sources in our catalogues should be of the same
order as a hypothetical age difference between disc-bearing solar type
and low-mass stars. We therefore perform a spatial analysis of the
distribution of class II and class III objects in each of our studied
clusters to verify that our method is indeed able to pick up the
signal in the average separations. 

The results are shown in Table~3, where it is clear that in all cases,
aside from Tr~37, the mean separation of disc-less
stars is larger than the mean separation of disc-bearing
sources. This demonstrates that the method we propose here is indeed
capable of picking up age difference of the order a few Myr in these
clusters. The signal is clearly present even though we included all sources
irrespective of spectral type. This latter
point further suggests that the timescales for dispersal cannot be a
strong function of spectral type. 

Tr~37 is the only outlier in our sample, where the disc-bearing objects
appear to be more spread out than the disc-less stars. However Tr~37
is known to show a distribution of ages between 
1 and 8 Myr with an age gradient through the cluster from east (older)
to west (younger). We have not included the so-called ``globule'',
where evidence for a new episode of star formation was reported by
Sicilia-Aguilar et al (2006 , see also Barentsen et
  al. 2011), however it is likely that the complex
star formation history of this region may be washing out the
signal in the spatial distribution. We therefore exclude Tr~37 from
further analysis.  

\begin{table}
\begin{center}
\begin{tabular}{lccccc}
\hline
Region    & \multicolumn{2}{|c|}{Selected Number}   &\multicolumn{2}{|c|}{d$_{av}$ [pc]}\\
               & ClassIII & ClassII &  ClassIII & ClassII \\  
Serpens      & 129   & 38  &  0.10 & 0.065 \\ 
Lupus III    & 64     & 55   &  0.11 & 0.077 \\
Taurus       & 181  & 121 &   0.55 & 0.38 \\
IC 348        &  93  & 206 &   0.18  & 0.16 \\ 
Cha I          & 99    & 89   &  0.12  & 0.05 \\ 
ChaII          & 40    & 19   &  0.21 & 0.12 \\
Tr 37$^*$      & 62    & 63 &  0.03  & 0.06   & \\
\hline
\end{tabular}
\\\small{$^*$ Excluding the Tr 37 West sources. See Text. }
\caption{The spatial distribution of the disc-bearing (Class II)
  versus disc-less (Class III) stars.}
\end{center}
\end{table}

\subsubsection{Test 2: Simulated clusters}
 \begin{figure*}
 \begin{center}
 \includegraphics[width=8cm]{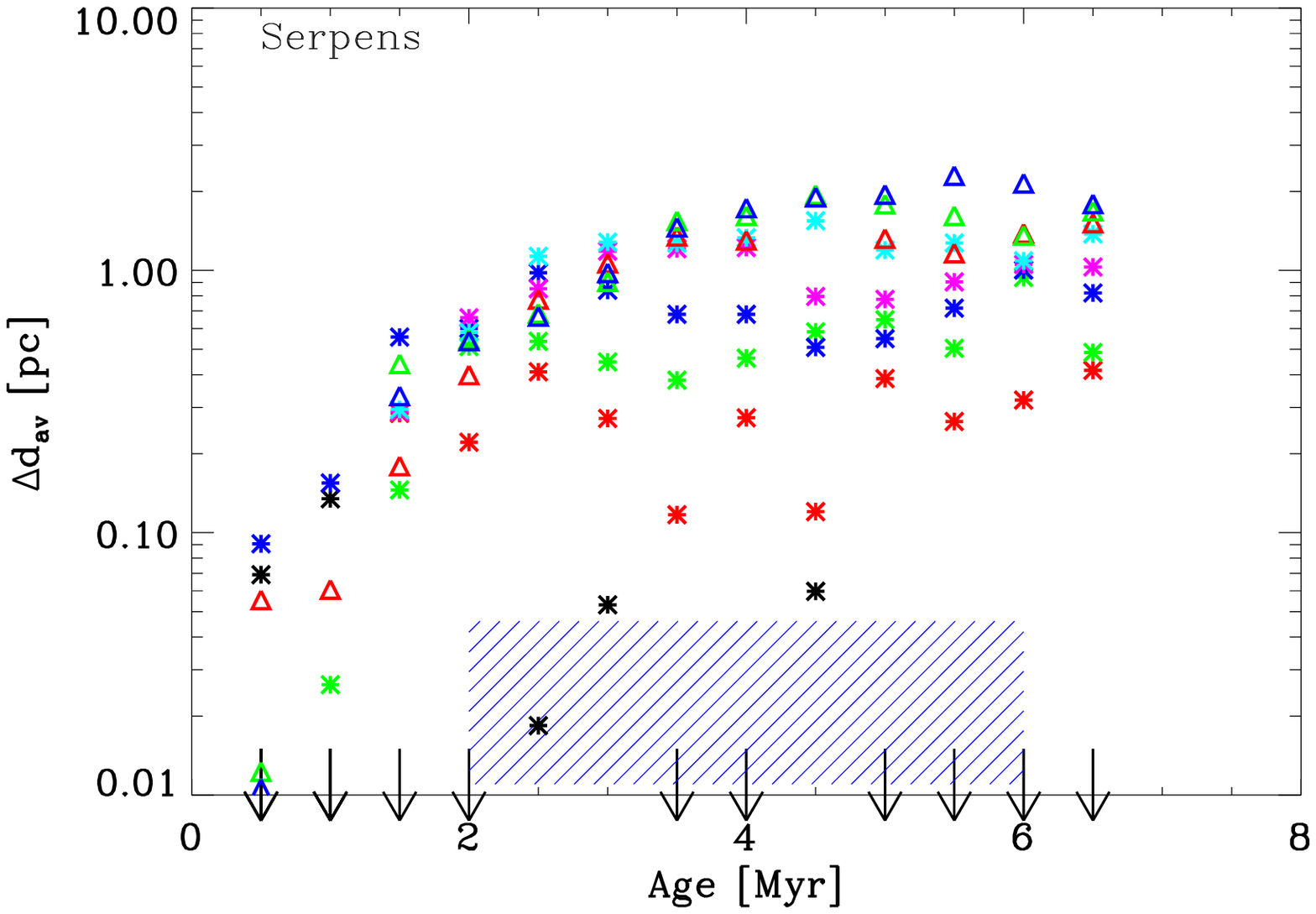}
 \includegraphics[width=8cm]{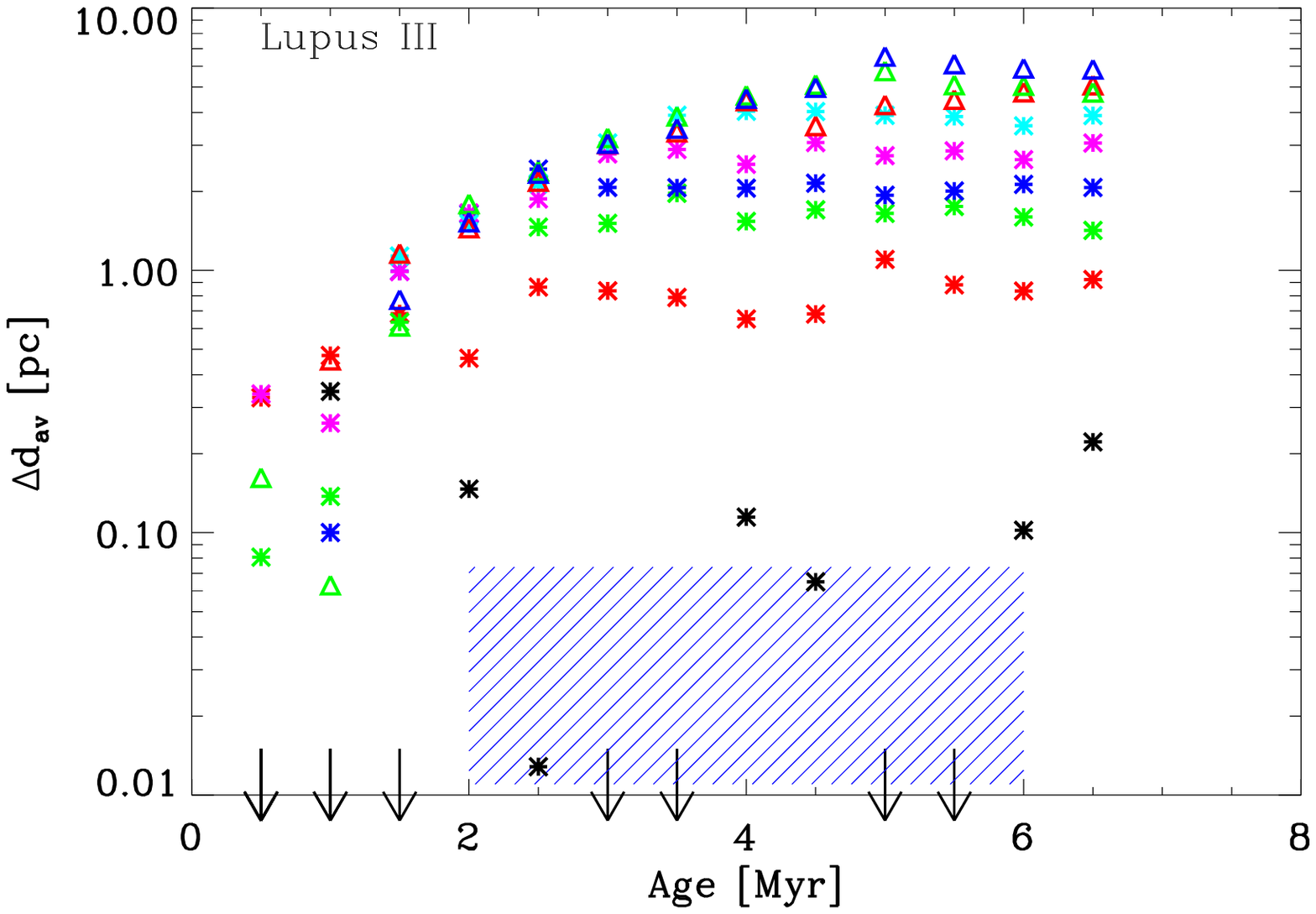}
 \includegraphics[width=8cm]{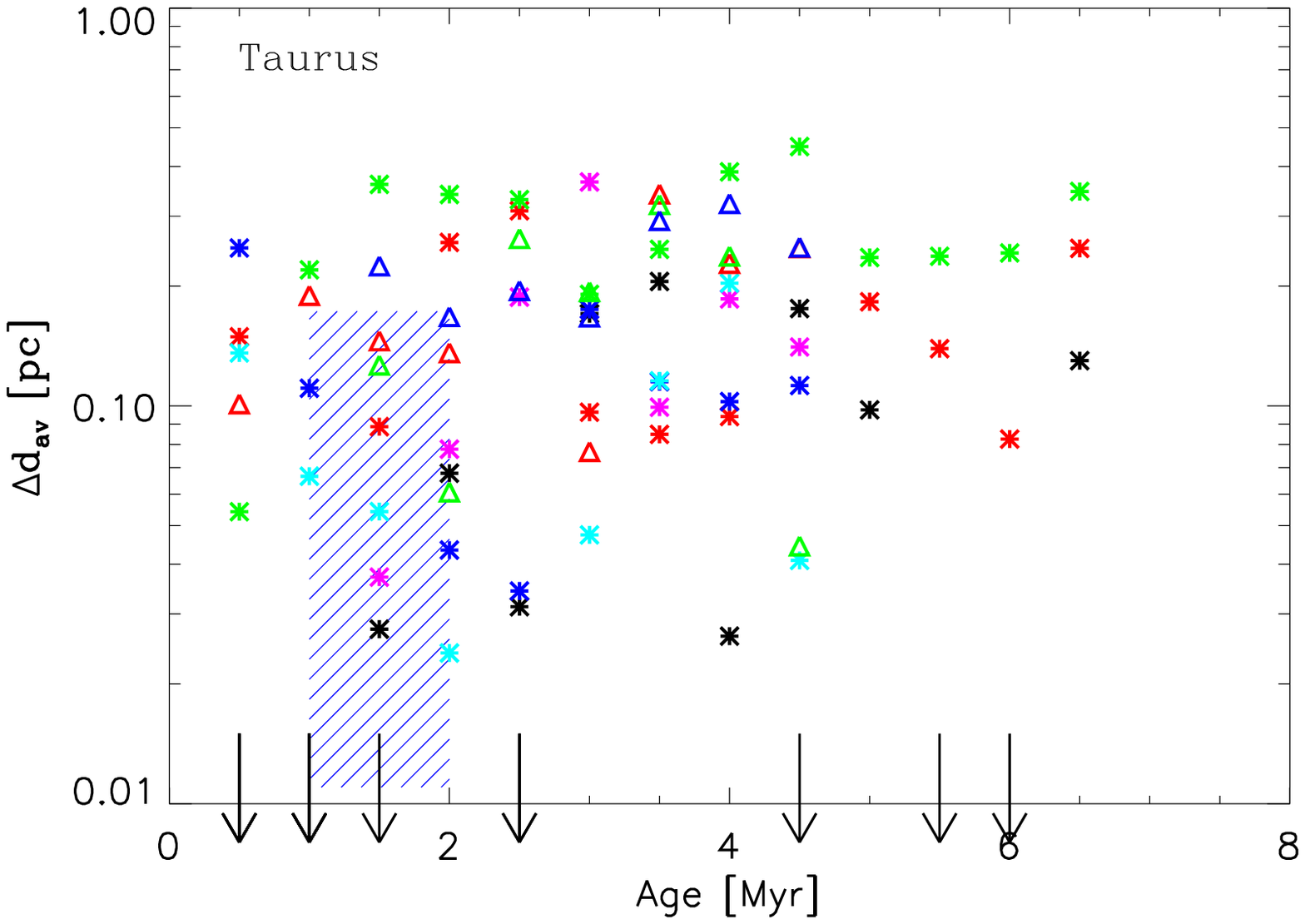}
 \includegraphics[width=8cm]{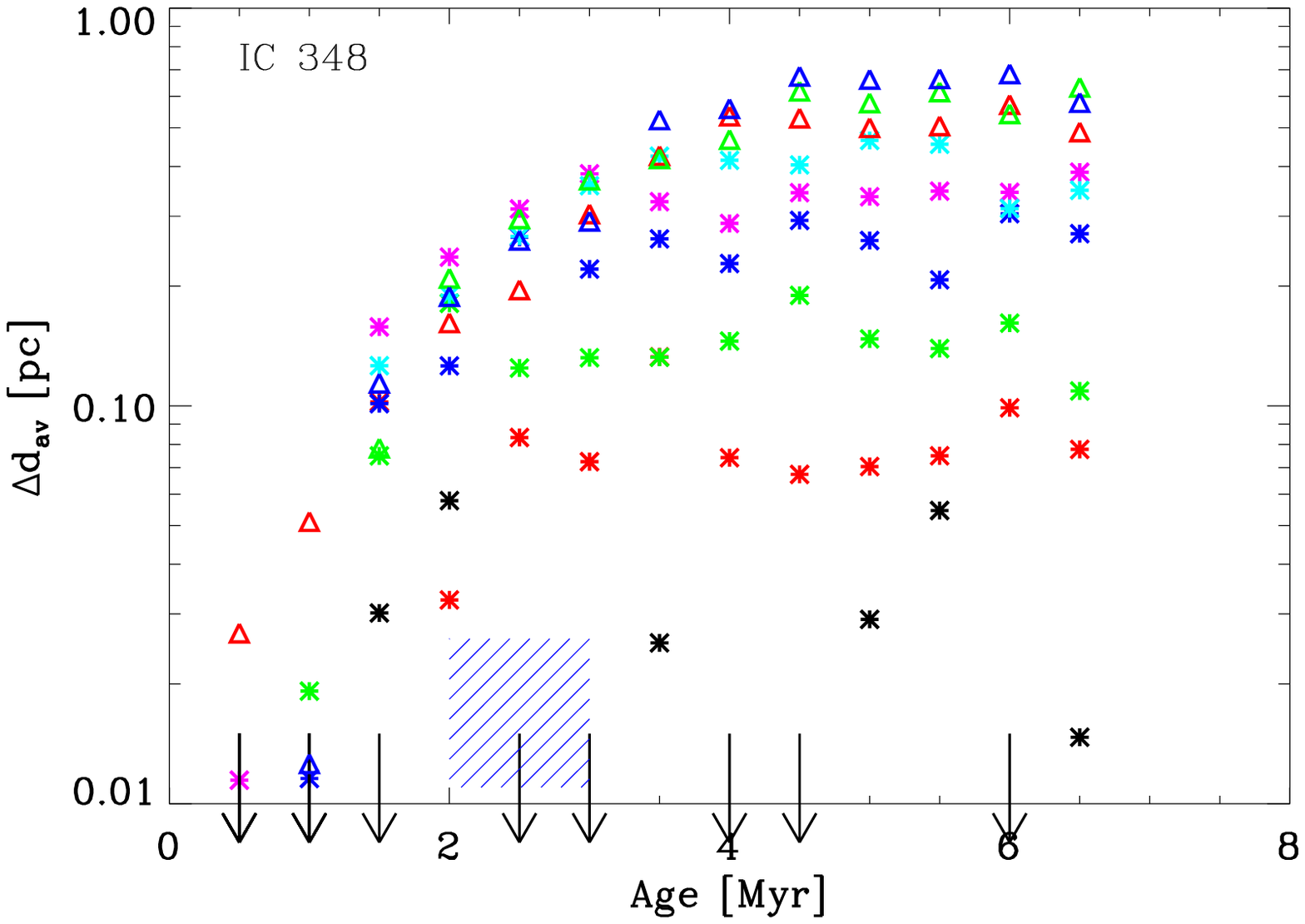}
 \includegraphics[width=8cm]{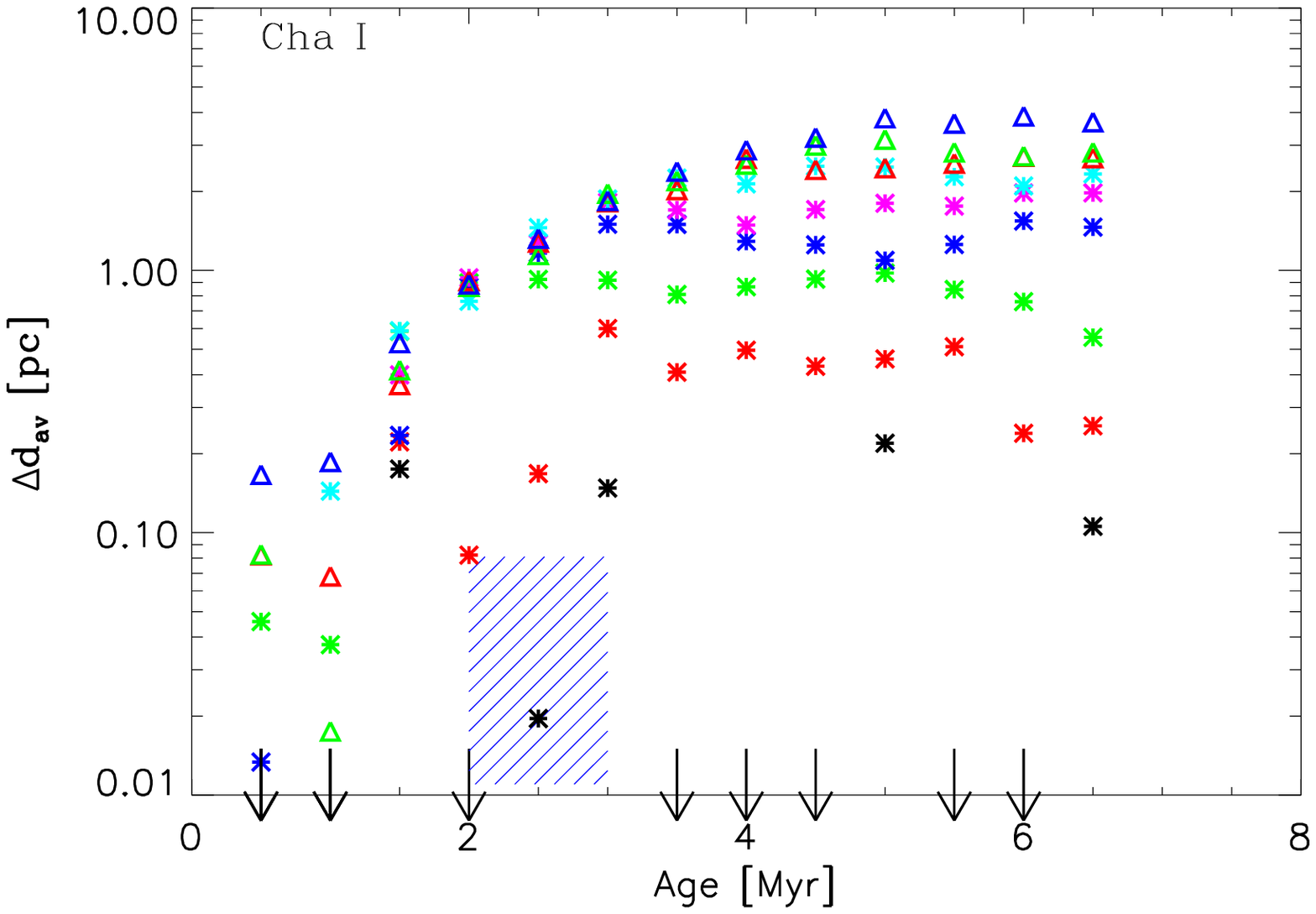}
 \includegraphics[width=8cm]{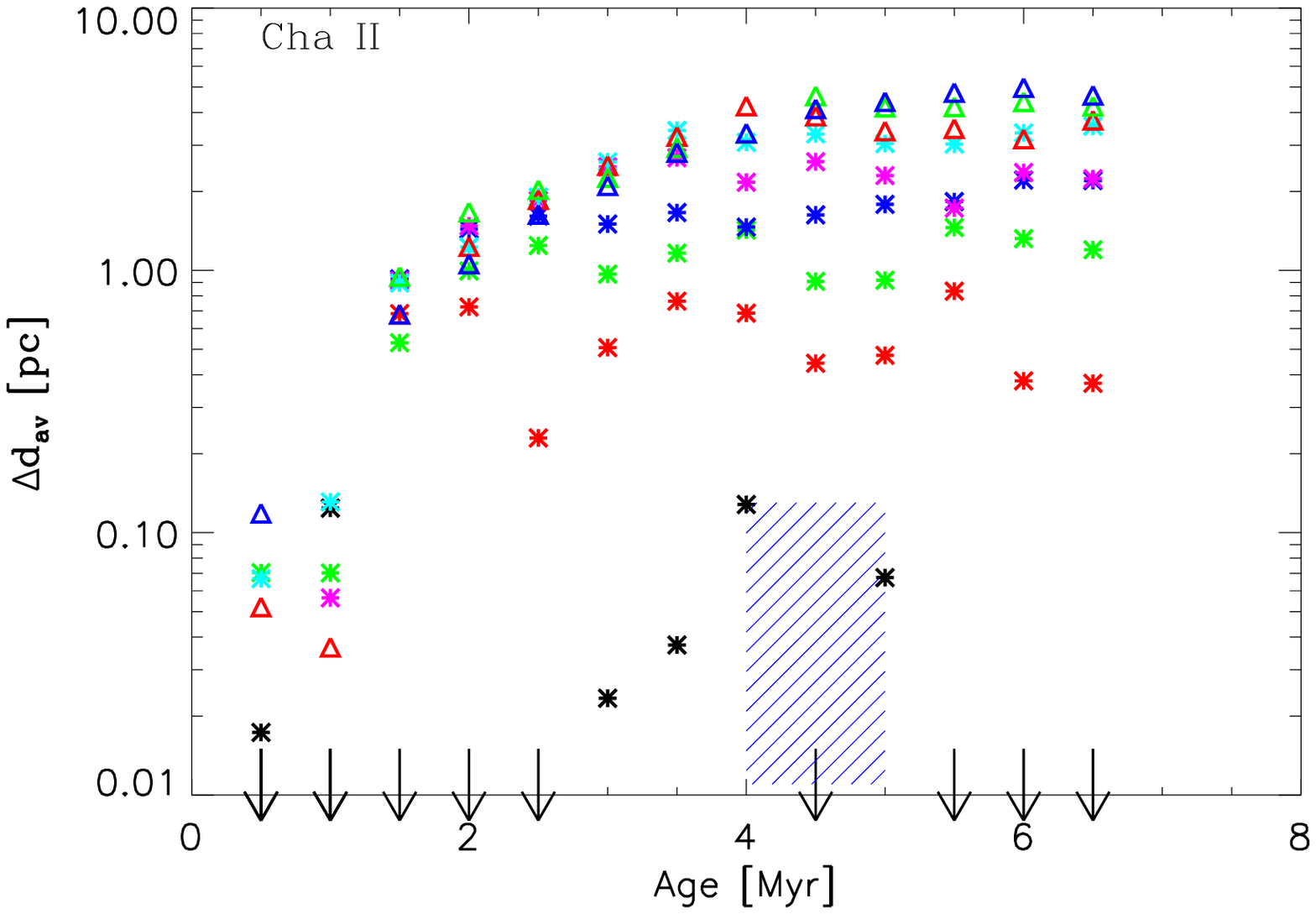}
\caption{Difference in nearest neighbour distance between the low mass
  and solar-mass stars as a function of age for simulated clusters
  with parameters as given in Table 1. The black, red, green, blue,
  magenta and cyan asterisks represent models with low-mass star disc lifetimes of 2, 3,
  4, 5, 6 and 7 Myr, respectively, while the black, red and green
  triangles represent models with low-mass star discs dispersal timescales of 8, 9 and 10 Myr. The
  solar type disc dispersal timescales are kept fixed at 2 Myr. The
  down pointing arrows represent negative values, or values smaller than the lower limit on the y-axis,  which have not been
  plotten on our log plot. The blue hashed box gives the 3 times Monte
  Carlo error obtained from the stochastic culling of the larger
  population in the spatial analysis of the observed regions (see
  section 2.1). } 
 \end{center}
 \end{figure*}

In order to
predict whether a detectable signal (above the three sigma error given
by the Monte Carlo culling described above) is to be expected, we have
performed the same MST-based spatial analysis on synthetic clusters of
sizes and velocity dispersions appropriate for each of the regions
listed in Table~1. We populate each cluster assuming that star formation is continuous
during the evolutionary timescale of the low-mass YSO discs and stars are
`born' in a three-dimensional fractal distribution. The simulations
presented here used a 3D fractal dimension of 1.7. Cartwright \&
Whitworth (2004) estimated fractal dimensions in the range 1.7
(e.g. Taurus) to 2.3 (e.g. Chamaleon). As a comparison the fractal
dimension of the Interstellar medium is thought to be 2.3 (Elmegreen
\& Falgarone 1996). In general the fractal dimension in young clusters
is expected to start off low (clumpy) and increase with age (e.g. Bastian et al 2009, 2011). We thus
use a low fractal dimension to assign the birthplaces of the YSOs. A
higher fractal dimension would of course weaken the expected signal,
so we have also investigated models with higher (2.3) fractal
dimensions and found that this only affects the results $\Delta
d_{av}$ by 25\% at most. $\Delta
d_{av}$ is defined as the difference between the mean $d_{av}$ of
disc-bearing low mass and solar-type stars.

The solar-type star discs are assumed to disperse after
2~Myr and we vary the lifetime of the low mass star discs, $\tau_{lm}$,
between 2 and 10
Myr in steps of 1 Myr. We assign the direction of motion of each star
stochastically with a velocity module equal to the literature value
of the velocity
dispersion measured for the given region (Table 1). 
For each simulation we measure $\Delta
d_{av}$, after stochastically culling the synthetic clusters to the
same number of sources available in the observed clusters (see Table
2).  We summarise the results of our Monte Carlo simulations in
Figure~1, where each region-specific simulation is plotted in
individual panels. The black, red, green, blue,
  magenta and cyan asterisks represent models with low-mass star disc lifetimes of 2, 3,
  4, 5, 6 and 7 Myr, respectively, while the black, red and green
  triangles represent models with low-mass star discs dispersal timescales of 8, 9 and 10 Myr. The
  solar type disc dispersal timescales are kept fixed at 2 Myr. We note that in our very simplified model, what matters is mainly the difference in the assumed mean disc dispersal timescale of solar types versus low mass, rather than the absolute ages. Here we explore parameter space by choosing to keep the solar type discs at the same value and vary the dispersal timescale of low mass discs to cover a difference in age from 0 to 8~Myr. This is a very simple statistical test and it cannot be used to suggest absolute disc dispersal timescales. This test has the only aim, in combination with the spatial analysis presented in the previous section, to provide an indication of an upper limit to the difference in disc dispersal timescales between solar type and low mass stars that is compatible with the null result found in the previous section. 
  
  The down pointing arrows represent negative values, which have not been
  plotted on our log plot. The blue hashed boxes show the 3~$\sigma$
Monte Carlo error obtained from the stochastic sampling of the
observed clusters. Figure 1 shows that for all clusters, apart from
Taurus the observations are not consistent for differences in the disc
lifetimes between low mass and solar type stars in excess of 1 Myr,
and are instead consistent with the two populations having basically
indistinguishable lifetimes. This result will be further discussed in
the context of the observations in the next section.

As mentioned above the scope of this very simplified test is only to
quantify the age differences to which the spatial distribution
analysis presented in the previous section is sensitive to, and to
this aim the age difference between the two population, rather than
the absolute dispersal age of each, is the important
parameter. However it should be also noted that the ``base age'',
i.e. the dispersal timescale of the shortest lived population, plays a
role for those regions with ages comparable to it. To address the
concern that the assumption of an older dispersal age for solar type
stars may weaken the expected signal in the younger regions, we have
repeated the experiment for the synthetic IC348 and ChaI models,
setting the dispersal timescales for solar-type discs to 3~Myr and to 4~Myr. We
show the results in Figure~3, where it is clear that our conclusions
are unaffected by the choice of the longer timescales even for these
young regions. Figure~3 still indicates that the null signal is still only consistent with 
differences smaller than 1 Myr in the mean disc dispersal timescale of discs around the two populations. 

 \begin{figure*}
 \begin{center}
 \includegraphics[width=8cm]{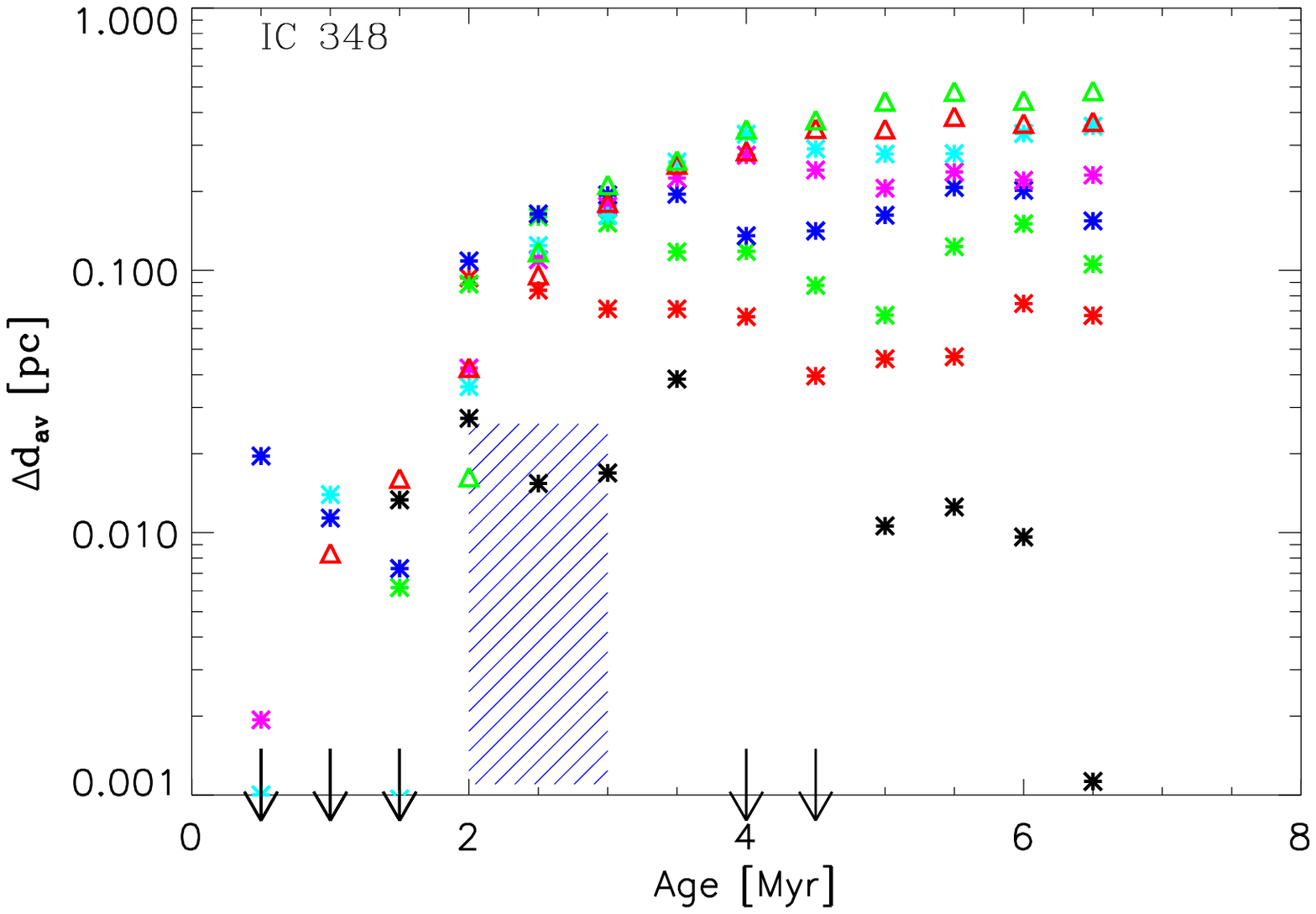}
 \includegraphics[width=8cm]{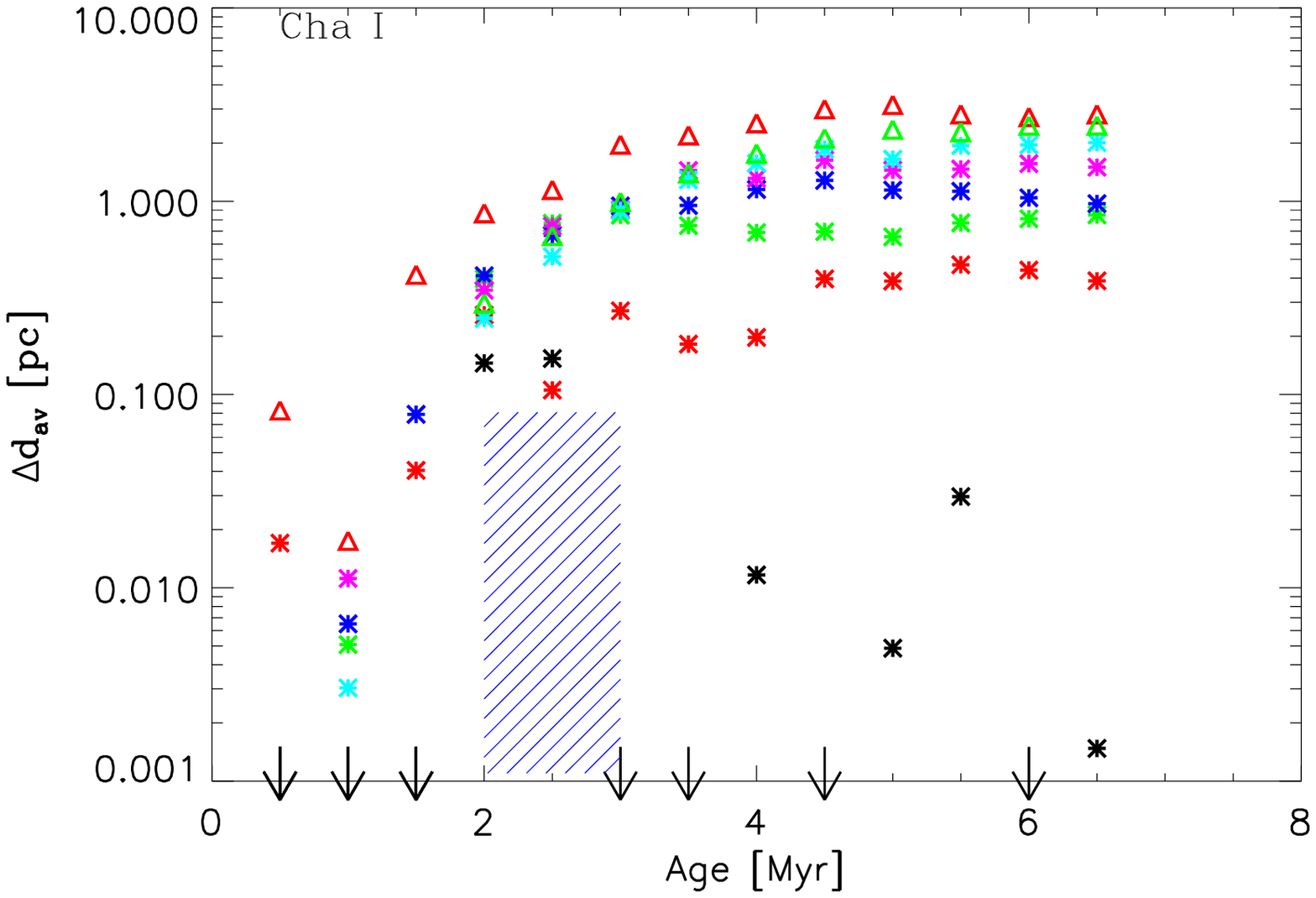}
  \includegraphics[width=8cm]{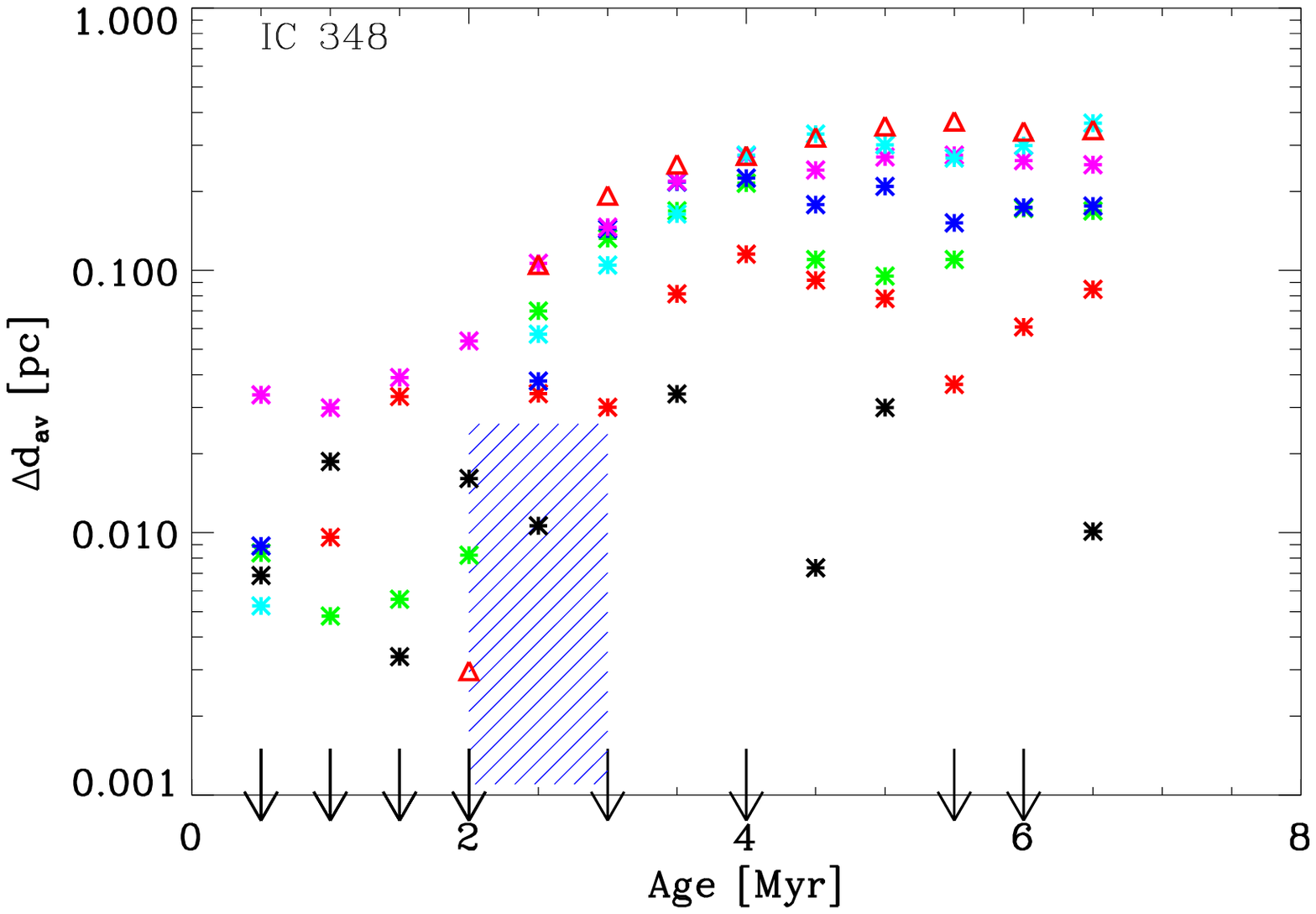}
 \includegraphics[width=8cm]{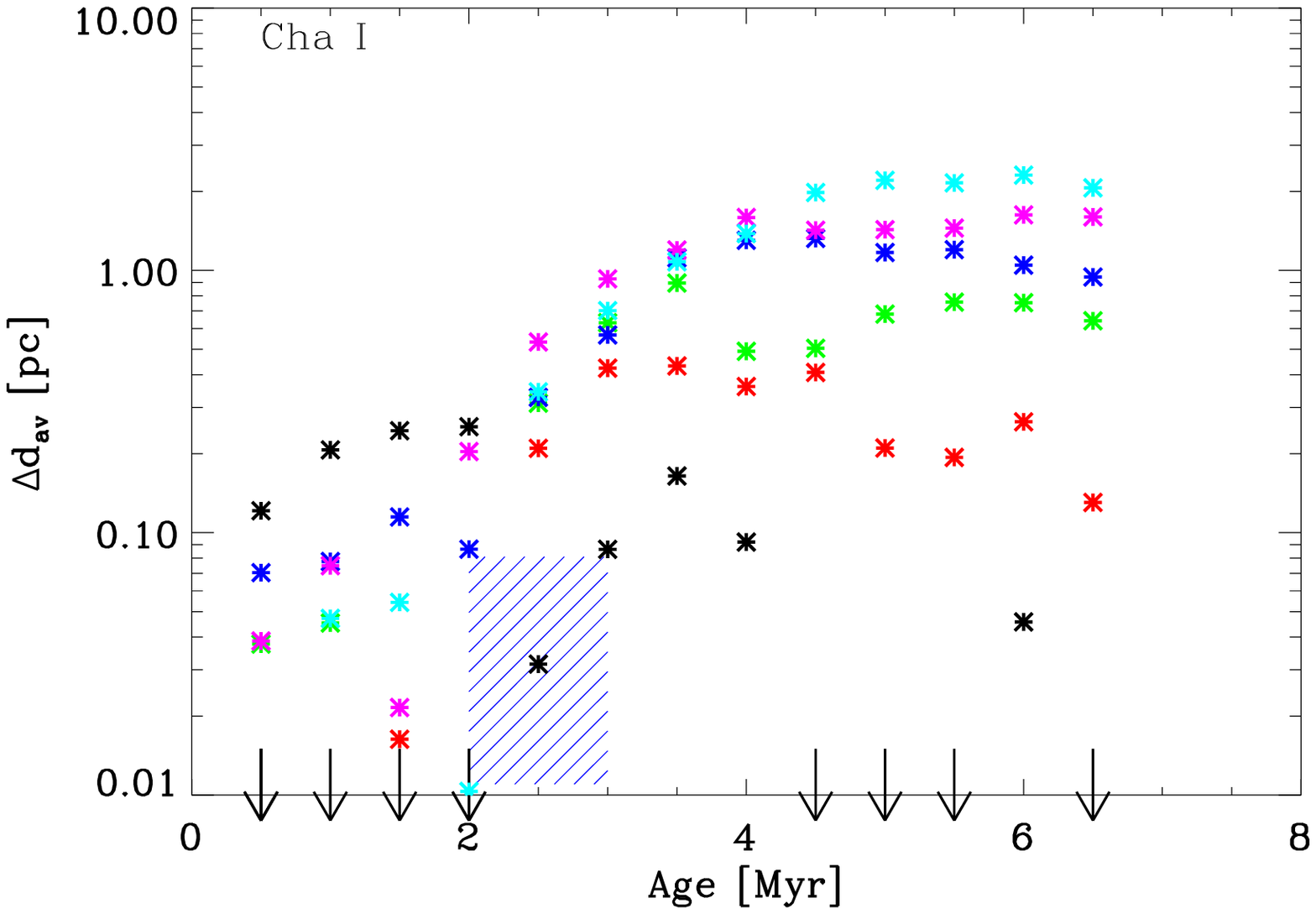}
\caption{Difference in nearest neighbour distance between the low mass
  and solar-mass stars as a function of age for simulated clusters
  with parameters as given in Table 1 for IC348 and ChaI. The black, red, green, blue,
  magenta and cyan asterisks represent models with low-mass star disc lifetimes of 2, 3,
  4, 5, 6 and 7 Myr, respectively, while the black, red and green
  triangles represent models with low-mass star discs dispersal timescales of 8, 9 and 10 Myr. The
  solar type disc dispersal timescales are kept fixed at 3 Myr (top panels) and 4 Myr (bottom panels). The
  down pointing arrows represent negative values (or values smaller than the y-axes limit, which have not been
  plotted on our log plot. The blue hashed box gives the 3 times Monte
  Carlo error obtained from the stochastic culling of the larger
  population in the spatial analysis of the observed regions (see
  section 2.1). } 
 \end{center}
 \end{figure*}

We note that, in theory, mass segregation could also result in lower
average nearest neighbour distances for the higher mass star. This
effect could easily be corrected for by comparing the distribution of
all YSOs (disc-bearing and disc-less). However, mass segregation is
not expected to be a dominant effect in the low mass low number
regions considered here, and, in any case, as will be further
discussed in the next session, we do not find any difference
in the distribution of YSOs as a function of mass (in the mass range
considered here), regardless of
whether they have an infra-red excess. Dynamical expulsion of low-mass
members from clusters could also be a problem by decreasing the
average nearest neighbour distance of the low-mass members. Again,
this effect is only expected to be relevant in a `classical' cluster,
i.e. a centrally concentrated dense object where dynamical interactions
between members are common. The star forming regions considered here
are all relatively low surface density objects (c.f.~Bressert et
al. 2010). Finally, bound clusters wold also evolve differently, of course,
from the simple description of expansion in our toy models, N-body
simulations are required to model these effect in 
detail and will be explored in future work; this is, however, not
relevant for the young low density regions studied here. 

\section{Results}

In Table~2 we summarise the results for the relative average
separations of low and solar-mass
disc-bearing stars in the seven regions studied. The difference
between the average nearest
neighbour distances for these two populations
($\Delta d_{av}$) is indicated. The errors refer to 
the uncertainties introduced by our stochastic culling of the more
numerous population, described in the previous section, and it represents 
the Monte Carlo error intrinsic to this procedure. In five out of the
six regions studied in this work (we exclude Tr 37 from the
discussion, see Section 2.2.1) the difference in $d_{av}$
between the low and solar-mass stars is always smaller than the three
sigma error from the Monte Carlo culling. The only exception is the
LupusIII region where we find a signal at the seven sigma level, but
with a negative sign (i.e. indicating smaller average distances for M-stars)
We  
note however that the average distances were computed using only 15
stars in each population and are distributed over an extended
non-spherical region. The same analysis performed on only the central
higher density region shows a much weaker but still detectable
negative signal
(roughly at the 4 
sigma level), but the low number statistics weaken any conclusions
from this one cluster alone. 

This null result can be interpreted in the contexts of the statistical
tests performed in Section 2.
 Our comparison of the spatial distributions of disc-bearing (class
 II) against disc-less (class III) 
objects, described in Section 2.2.1, showed that our
method is able to recognise an age difference between 
these two classes of objects. This suggests in the first instance that
if a similar age difference existed between disc-bearing low mass
objects and disc bearing solar type objecst, it should also be picked
up by the same method. Our Monte Carlo simulations of synthetic
clusters, discussed in Section 2.2.2, allows us to determine limits on
the maximum average age difference between low mass and solar type
discs in each of the cluster. Figure 1 shows that for all regions,
except for the young Taurus region, our results are consistent with no age difference
between the two populations. 

The simulated cluster analysis presented in the previous sections
predicts that for an age difference between the low-mass and
solar-type disc population greater or equal to 1 Myr the $\Delta
d_{av}$ should be much larger than the 3 $\sigma$ upper limits given
by the Monte Carlo errors (blue hashed boxes in Figure 1), thus arguing for
similar dispersal timescales (black asterisks in Figure 1) for both
populations. 
Taurus is the only exception, where the non-detection of a signal
cannot be interpreted as the two populations being coeval. This result
is largely driven by the extremely young age of the region, its large
physical extension and low velocity dispersion which conspire to give
an undetectable signal. 

On the basis of these considerations we conclude that there is no evidence of a significant difference in the lifetimes of discs around low-mass and solar-type stars on the basis of their relative spatial distributions. 
Indeed, the statistically indistinguishable spatial distribution of
the low-mass and
solar mass stars in Serpens, Lupus III, ChaI, ChaII and IC 348 argues
against the hypothesis that discs around low mass stars live significantly
longer than discs around solar-mass stars, although we
cannot rule out differences of less than 1 Myr (see Section~2). 

%We note however that indistinguishable spatial distributions would also
%be expected if the intrinsic age spread of amongst all members
%of the regions we studied were smaller than roughly 1~Myr. 
%While our method cannot discriminate between the two scenarios, we
%find that a very narrow star formation period is probably unlikely,
%given that class I and class II YSOs in the same clusters have been
%shown to have distinct spatial distributions consistent with class II's
%being older that class I's and thus more spread out (e.g. Gutermuth
%et al 2008). The very narrow age spread required to   
%wash out the signal in our experiment would be expected to do the same
%in that case. 

To summarise, our simple spatial analysis using MSTs provides
evidence against significantly longer disc dispersal timescales around late-type stars
(e.g. M-dwarf) compared to their solar-mass counterparts. Low number
statistics prevents us from providing tight constraints on the allowed
difference in disc survival timescales in these two populations, but
our analysis, based on simple models, indicates that it is consistent with similar disc
dispersal timescales, with differences certainly smaller than the
average age difference between CTTs and WTTs. 

\section{Acknowledgments}
We thank the referee for a constructive report that helped us improve
our paper significantly. We thank Cathie Clarke for useful comments with regards the statistical analysis. We thank Aurora Sicilia-Aguilar for providing useful information on the age distribution in the Tr 37 Cluster.

\end{document}